\documentclass[preprint,showpacs,preprintnumbers,amsmath,amssymb,endfloats*]{revtex4}
\usepackage{graphicx}

\begin{document}
\def \ee {\varepsilon}
\thispagestyle{empty}
\title{
Comment on ``Analytical and numerical verification of the
Nernst heat theorem for metals''
}

\author{
G.~L.~Klimchitskaya${}^1$
and V.~M.~Mostepanenko,${}^2$
}

\affiliation{
${}^1$North-West Technical University, Millionnaya Street 5,
St.Petersburg, 191065, Russia\\
${}^2$Noncommercial Partnership  ``Scientific\\ Instruments'', 
Tverskaya Street 11, Moscow, 103905, Russia 
}

\begin{abstract}
Recently, H{\o}ye, Brevik, Ellingsen and Aarseth
(quant-ph/0703174) claimed that the use of the Drude dielectric
function leads to zero Casimir entropy at zero temperature
in accordance with Nernst's theorem. We demonstrate that their
proof is not applicable to metals with perfect crystal
lattices having no impurities. Thus there is no any contradiction
with previous results in the literature proving that the
Drude dielectric function violates the Nernst theorem for the Casimir 
entropy in the case of perfect crystal lattices. We also indicate
mistakes in the coefficients of their asymptotic expressions
for metals with impurities.
\end{abstract}
\maketitle

As correctly mentioned in the Introduction of Ref.~\cite{1},
the relaxation frequency of a metal $\nu(T)$ goes to zero
 when temperature $T$ goes to zero. In accordance with
the Bloch-Gr\"{u}neisen law at low temperatures
$\nu(T)\sim T^5$.  It should be
particularly emphasized that the Bloch-Gr\"{u}neisen law
is established for metals with perfect crystal lattices
having no impurities. This law is also valid for metals
with impurities at temperatures larger than 3--4\,K.
Although real metals have some small fraction of impurities, the
model of perfect crystal lattice is basic in all theoretical
condensed matter physics. Many important results in this
field, including the theory of electron-phonon interactions,
are obtained for perfect crystal lattices. The Casimir
entropy in the case of a perfect crystal lattice, if calculated 
correctly, must satisfy the Nernst theorem and all other
requirements of thermodynamics. The reason is that the
perfect crystal lattice is a truly equilibrium system
with a nondegenerate dynamical state of lowest energy.
Consequently, in accordance with quantum statistical
physics, the entropy at zero temperature must be equal to zero
\cite{2}.

The analytical derivation of the thermal correction to the
Casimir energy between two Au plates in Ref.~\cite{1}
is based on the Drude model and uses the condition
\begin{equation}
\zeta_m(T)\ll\nu(T).
\label{eq1}
\end{equation}
\noindent
This condition should be satisfied by sufficient number of
Matsubara frequencies $\zeta_m$ with $m=1,\,2,\,3,\ldots$
[see Eqs.(5) and (9) in Ref.~\cite{1}; note that
Ref.~\cite{1} omits the lower index $m$ and the argument $T$].
Here $\zeta_m(T)=2\pi kmT/\hbar$, $k$ is the Boltzmann
constant and $m=0,\,1,\,2,\ldots\,$.

It is easily seen that in the case of perfect crystal lattice the
condition (\ref{eq1}) does not hold for any nonzero Matsubara
frequency. In fact, according to Ref.~\cite{1}, for Au
$\nu(T=300\,\mbox{K})=34.5\,$meV whereas
$\zeta_1(T=300\,\mbox{K})=161.9\,$meV. Thus 
$\nu(T=300\,\mbox{K})<\zeta_1(T=300\,\mbox{K})$ in contradiction with
assumption (\ref{eq1}).
Taking into account that $\zeta_m=m\zeta_1$, the same inequality
is valid for all nonzero Matsubara frequencies.
When $T$ decreases from room temperature up to approximately
$T_D/4$, where $T_D$ is the Debye temperature ($T_D=165\,$K
for Au \cite{3}), $\nu(T)\sim T$, i.e., decreases with decreasing
temperature following the same law as $\zeta_m$.
This preserves the inequality
\begin{equation}
\nu(T)<\zeta_m(T),\quad m=1,\,2,\,3,\ldots\,.
\label{eq2}
\end{equation}
\noindent
At $T<T_D/4$ the relaxation frequency decreases even more rapidly 
than $\zeta_m$ with decreasing $T$ (i.e., as $\sim T^5$ according 
to the Bloch-Gr\"{u}neisen law due to electron-phonon
collisions and as $\sim T^2$ at liquid helium temperatures due
to electron-electron scattering). As a result, with the decrease
of temperature for perfect crystal lattices it holds
\begin{equation}
\nu(T)\ll\zeta_m(T),\quad m=1,\,2,\,3,\ldots\,.
\label{eq3}
\end{equation}

This inequality is just the opposite of the inequality (\ref{eq1}) 
used in the derivation of Ref.~\cite{1}. Thus, all the results, 
obtained in Ref.~\cite{1}, are inapplicable to perfect crystal
lattices. According to Ref.~\cite{1} ``the Nernst theorem is
not violated when using the realistic Drude dispersion model''
and this conclusion ``is clearly in contrast to that presented
in various works \cite{4,5,6,7,8}''. These formulations are,
however, misleading. References \cite{4,5,6,7,8} deal with
perfect crystal lattices and prove that for such lattices the use 
of the Drude model leads to the violation of the Nernst heat theorem.
As explained above, the derivation in Ref.~\cite{1} is not
applicable to perfect crystal lattices because it uses the
inequality (\ref{eq1}) which is just the opposite to the
inequality (\ref{eq3}) satisfied for perfect lattices.
Thus, there is no contradiction between the results of Ref.~\cite{1}
and Refs.~\cite{4,5,6,7,8}. Note that all above explanations concerning 
the inequalities (\ref{eq1})--(\ref{eq3}) are contained in Ref.~\cite{4}.
However, they were simply ignored in Ref.~\cite{1}.

What is in fact found in Ref.~\cite{1} [see Eq.(31)]
is the analytic behavior of the low-temperature thermal correction
to the Casimir energy using the Drude model for crystal lattices
with impurities:
\begin{equation}
\Delta F=C_1T^2(1-C_2T^{1/2}+\ldots),
\label{eq4}
\end{equation}
\noindent
where $C_1$ and $C_2$ are constants.
According to this correction, at very low temperatures the Casimir 
entropy abruptly jumps to zero starting from negative values. 
Thus formally the Nernst heat theorem is satisfied when impurities
are present. This result is not new. It was first found 
by B.E.\ Sernelius in Ref.~\cite{9} and has been acknowledged in
Refs.~\cite{4,5,6,7,8}. Previously this result was proven
only numerically. Reference~\cite{1} provides an analytical
proof.

However, the values of numerical coefficients $C_1$ and $C_2$
for Au in Eq.~(\ref{eq4}) are determined in Ref.~\cite{1}
incorrectly. To calculate these coefficients, Ref.~\cite{1}
uses the Au relaxation frequency $\nu(T=300\,\mbox{K})=34.5\,$meV
and the inequality (\ref{eq1}).
However, as explained above, at room temperature and also at 
much lower temperatures in the application range of the
Bloch-Gr\"{u}neisen law, the inequality (\ref{eq1}) 
 is violated and exactly the opposite
inequality (\ref{eq3}) is valid. The inequality (\ref{eq1})
used in Ref.~\cite{1} becomes valid only for imperfect
lattices at very low $T$ when, due to the presence of
impurities, the relaxation frequency deviates from the
Bloch-Gr\"{u}neisen law and takes a nonzero $T$-independent
residual value $\nu_0$. For typical Au samples the residual
relaxation frequency is approximately equal to
$\nu_0\approx 34.5\times 10^{-3}\,$meV, and for the best 
samples which are most close to the perfect crystal it
is even 3 orders of magnitude lower:
$\nu_0\approx 34.5\times 10^{-6}\,$meV \cite{3}.
In order that at least the first 10 Matsubara frequencies 
satisfy the inequality
\begin{equation}
\zeta_m(T)\ll\nu_0,
\label{eq5}
\end{equation}
\noindent
the temperature must be $T<10^{-3}\,$K for typical Au samples
and $T<10^{-6}\,$K for the best Au samples. For the applicability 
of asymptotic expression (\ref{eq4}) [Eq.(31) in Ref.~\cite{1}]
the temperatures must be additionally at least one order of
magnitude less.

As was mentioned above, to calculate the values of the
coefficients $C_1$ and $C_2$ Ref.~\cite{1} uses the
value $\nu(T=300\,\mbox{K})=34.5\,$meV. The correct
values to be used instead are $\nu_0=34.5\times 10^{-3}\,$meV
for typical Au samples and $\nu_0=34.5\times 10^{-6}\,$meV
for the best Au samples. As a result, Eqs.(13), (18) and (30) in
Ref.~\cite{1} lead to the following values of coefficients
in Eq.~(\ref{eq4}):
\begin{eqnarray}
&&
C_1=5.81\times 10^{-10}\,\mbox{J/(m${}^2$\,K${}^2$)},
\quad
C_2=95.75\,\mbox{K${}^{-1/2}$}
\quad
\mbox{(typical Au samples)},
\nonumber \\
&&
C_1=5.81\times 10^{-7}\,\mbox{J/(m${}^2$\,K${}^2$)},
\quad
C_2=3028.0\,\mbox{K${}^{-1/2}$}
\quad
\mbox{(best Au samples)}.
\label{eq6}
\end{eqnarray}
\noindent
This should be compared with the values presented in
Ref.~\cite{1}:
\begin{equation}
C_1=5.81\times 10^{-13}\,\mbox{J/(m${}^2$\,K${}^2$)},
\quad
C_2=3.03\,\mbox{K${}^{-1/2}$}.
\label{eq7}
\end{equation}

The results of numerical computations in Ref.~\cite{1} were found
to be in agreement with the asymptotic expression (\ref{eq4})
containing the wrong coefficients (\ref{eq7}). The reason is that
in numerical computations the room temperature relaxation frequency
$\nu(T=300\,\mbox{K})=34.5\,$meV was also used incorrectly 
within the wide temperature region from 0.01\,K to 800\,K.
To obtain the correct computation results, from 4--5\,K to 800\.K
the actual temperature dependence of the relaxation frequency
on $T$ should be employed (given by the Bloch-Gr\"{u}neisen law
and the linear dependence). For temperatures around zero the
residual relaxation frequency for Au, $\nu_0$, depending on the
concentration of impurities, must be applied.

To conclude, Ref.~\cite{1} finds (up to incorrectly determined
coefficients) the low-temperature behavior of the thermal correction 
to the Casimir energy in the configuration of two Au plates with
impurities, using the permittivity of the Drude model. 
Although the results of Ref.~\cite{1} are in formal agreement
with the Nernst theorem, there is no contradiction with
the results of Refs.~\cite{4,5,6,7,8} demonstrating the
violation of the Nernst theorem in the Drude model approach
for perfect crystal lattices. The reason is that the condition
used by the authors of Ref.~\cite{1} in their derivation is
violated for perfect lattices and can be applied to only 
lattices with impurities.

It must be emphasized that the results of Ref.~\cite{1} do not
solve the problem of inconsistency of the Drude model with basic
thermodynamic principles in the application to the Casimir entropy,
as the authors claim. Reference~\cite{1} recognizes that
``a simple physical model of course cannot be permitted to
run into conflict with thermodynamics''. However, as is
clearly seen in the above and from Refs.~\cite{4,5,6,7,8},
the Drude model violates the Nernst heat theorem for the
Casimir entropy in the case of metals with
perfect crystal lattices. This alone makes the Drude model
approach to the Casimir force unacceptable as being in
contradiction with quantum statistical physics. According to
the authors of Ref.~\cite{1}, 
the approaches with nonzero contributions of
the transverse electric term at zero frequency (recall that
in the Drude model approach this term does not contribute
at $\zeta=0$) would violate the Nernst theorem.
This is misinformation. As is rigorously proved in
Refs.~\cite{4,10}, both the plasma model approach
and the impedance approach are in agreement with the
Nernst theorem, and both of them include
a nonzero contribution from the transverse electric term
at zero frequency. Thus, although for metals with
impurities the Drude model approach leads to zero Casimir entropy
at zero temperature, this approach is theoretically invalid
and fails to provide a consistent description of the
thermal Casimir force in the framework of the Lifshitz theory.

\section*{Acknowledgments}
The authors are grateful for helpful discussions with all
co-authors of our publications on this subject, i.e.,
V.~B.~Bezerra, M.\ Bordag, R.\ S.\ Decca, E.\ Fischbach, 
B.\ Geyer, D.\ E.\ Krause, D.\ L\'opez,  
U.\ Mohideen and C.\ Romero.

\end{document}